# Investigation of Audience Interaction Tools from the Perspective of Activity Theory

**Shadi Esnaashari**
School of Engineering and Advanced Technology
Massey University
Auckland, New Zealand
Email: S.Esnaasahri@massey.ac.nz

**Anuradha Mathrani**
School of Engineering and Advanced Technology
Massey University
Auckland, New Zealand
Email: A.S.Mathrani@massey.ac.nz

**Paul Watters**
School of Engineering and Advanced Technology
Massey University
Auckland, New Zealand
Email: P.A.Watters@massey.ac.nz

## Abstract

Maintaining engagement of large audiences is not easy. Traditionally, lectures and presentations have relied on one-way mmunication from the presenter to the listening audience. Without receiving ongoing feedback, speakers cannot be sure that their delivery is at an appropriate pace, or that their message is being received and understood by their audience. This study suggests using a real-time audience engagement solution (Xorro-Q) to facilitate synchronous interaction between lecturers and their student audiences. Using activity theory as a theoretical framework we conducted a study to investigate student participation and engagement with an audience interaction tool in two undergraduate computing courses. In one classroom setting, the lecturer employed continuous informal discussion-based teaching activities with Xorro-Q tool. The other classroom setting used Xorro-Q to formally assess students' subject knowledge by using traditional quiz type questions. The preliminary findings showed that audience participation tool has a promising direction for engaging students in both classroom settings. The class which adopted continuous informal discussion approach rendered more enjoyment among students, although the traditional formal assessment activities showed higher student participation. Thought these findings are at a very initial stage, they give some indication on how real-time audience engagement tools can be developed within classroom settings for assisting in teaching and learning practices.

## 1 INTRODUCTION

Today digital technologies are integrated with people's lives. These technologies have become a primary source of acquiring information for people. Therefore, it is conducive if these technologies can be combined in students' learning process in educational settings. One of the strategies to leverage the benefits with technologies in classes is the use of synchronous tools. In large classes, it is often hard to maintain participation of students. Students may prefer to stay silent in classes due to different reasons such as peer pressure, risk avoidance, anxiety, and cultural reasons. It is also mentioned that there is fear among students to interact with teaching staff in class due to evaluation anxiety, being judged by others, or being the focus of attention (Weaver and Qi 2005). However, in large audiences knowing the level of audience understanding of the current topic can aid lecturers in many ways. Lecturers can then adapt their teaching style and their speed of teaching to increase the understanding of the audience. Thus, getting quality and timely feedback can help teachers overcome the problem of disengagement and boredom in large audiences (Cue 1998; Zhu 2007).

This study investigates students' perception in regards to using audience interaction technologies to bring about synchronous interactions in classrooms. This has been done by using a web-based tool (Xorro-Q) in two classroom settings. A theoretical framework utilising Activity Theory (AT) underpins the research study design. Two different teaching pedagogies, namely activity-based (to stimulate discussions in classrooms) and traditional (to assess student learning) have been employed. The objective of the study was to understand the students' views on how the tool helped them in the process of learning and becoming engaged in the two separate classroom environments. Findings have indicated that audience-interaction tools such as Xorro-Q can assist in both traditional and activity-based pedagogies and give a promising research direction for enhancing student engagement in different classroom settings. The following sections give an overview on background literature and theoretical framework, study design, preliminary study findings and proposes future research directions.

## 2   LITERATURE REVIEW

To enhance the teaching and learning experience of students, it is important to increase student motivation and keep them engaged in class. Teachers use different techniques in class for their teaching. Many studies have been conducted about different styles of teaching in lectures (McKeachie 1990; Saroyan and Snell 1997). In traditional teaching, the teacher is central, as subject content is transferred one-way from teacher to the students. In this method of teaching, students receive the information from the teacher or from the textbook. Assessment is based on either right or wrong answers. The curriculum in this method encourages a lecturing type of teaching because there is a strong focus on facts which involve a large number of subject related vocabulary (Leonard and Chandler 2003, p. 5). However, new teaching methodologies emphasize collaborative learning and stress that learning can be improved through ongoing dialogue between teachers and learners (Draper et al. 2002). Students learn more by engaging in class and doing activities compared to sitting in class and passively listening to the lecturer. Collaborative learning accentuates the characteristics of the group rather than those of the individual (Chickering and Gamson 1987; McConnell 1996). The role of interactivity has a positive impact on the success of the course and consequently on the overall student learning process (Steinert and Snell 1999).

Sibley and Spiridonoff (2010) suggest using team and group learning in order to accelerate active and collaborative learning. Team based learning (TBL) suggests changing the class format from lecture-based to team-based. Team members use their time to apply and evaluate course materials instead of just acquiring the materials. Further, in TBL each member of the group is responsible for a specific portion of the assignment, which is considered better compared to group learning, since some students may not pull their part in the group. Weaver and Qi (2005) also support the idea that students learn more when they are actively engaged in the class. They suggest that teachers should use different methods to increase class participation for supporting student-driven learning.

A large number of empirical studies affirm that students learn better when they tackle questions in class rather than passively listen to answers (Waldrop 2015). Researchers from different disciplines emphasize the importance of using active learning in undergraduate classes to enhance student learning (Kober 2015; Singer and Smith 2013). This can be done through workshops, classroom discussions, debates, and using examples which are not taken directly from the textbook. Studies suggest that students' retain more subject knowledge and their scores could improve by 20% when active learning is used, whereby students engage in discussions during classes (Dörner 1996; Wieman 2014).

These findings resonate with a long established learning theory that is called activity theory (AT), which was first defined by Leont'ev (1974). AT was later extended by Engeström (2001), and it has now become a useful theoretical framework for exploring social relationships across different disciplines involving human computer interaction scenarios such as in requirement gathering, software development, education, and healthcare (Georg and France 2013; Hasan 1999). Further, Murphy and Manzanares (2008) add that each element of activity theory can be impacted by emerging technologies. For example, the component of traditional classrooms can be replaced by virtual classrooms and tools such as chalk and board could be replaced by emails, software apps and message texting. There are many studies conducted about the use of different technologies in educational environments (Park and Farag 2015; Ravishankar et al. 2014). This study too will explore classroom settings using an audience interaction tool (Xorro-Q) to identify ways to enhance student participation and learning. Xorro-Q is a new web-based tool that can be used to increase the level of engagement between students and lecturers. Xorro-Q supports a variety of pedagogies needed for assessing students' understanding, providing feedback to the students based on their learning, enabling classroom discussion for students and lecturers, and enabling lecturers to instantly adapt their instructions based on student responses.

## 3    THEORIZING WITH ACTIVITY THEORY

Vygotsky (1980) (published first 1931) and his collaborators in Russia coined the socio-cultural approaches to learning and development (Feryok 2012; Lantolf and Appel 1994). They argued that human mental functionalities are mediated processes organized by socio-cultural artifacts. Several socio-cultural theories have been derived from Vygotsky's work (Lantolf et al. 2000; Valsiner 2007; Van Lier 2002). Engeström (2001) state that all these approaches share one common theme that is human action is mediated. However, these approaches differ in how mediation is actually theorized. Leont'ev (1974) developed Vygotsky's meaningful social activity as a form of mediation which was known as activity theory. Leont'ev (1974) state that activity theory differentiates between individual goals and objectives at the action level, and social goals and objectives are differentiated at the activity level. Concrete operations are then utilized to achieve goals and objectives.

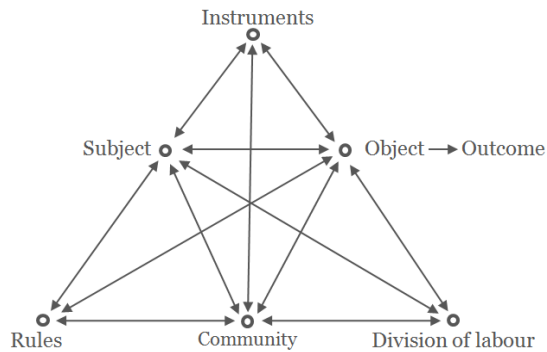

*Figure 1: Activity theory (Engeström, 2001)*

A model of activity theory is depicted in Figure 1. In the model the top triangle represents the mediating elements (tools, rule, and division of labour) while the inner triangle represents the mediated elements (subject, community, and object). The main primary focus of any activity is the object (or aim). The use of the term 'object' causes much confusion and some authors prefer to use the term 'aim' instead (Georg et al. 2015). The overall aim of the mediating relations is an outcome (or transformation) as a result of execution of the activity. Any activity is performed using combination of the subject, the object and the tool (instrument). The operation and action would affect the outcome. The subject would be individuals or group of actors involved in the activity. The object would be the production of the activity. Instrument would be anything used in the transformation process. The community would be the interdependent aggregate who share a set of social meanings. Rules are guidelines which guide the actions in order to be acceptable by the community. The division of labour would be the task specialization by individual members of the group.

Activity theory has been used in different disciplines in education and organizational learning (Basharina 2007; Foot 2001; Murphy and Rodriguez-Manzanares 2008).  This study has been theorized with AT contextualized elements to investigate student engagement with an audience interaction tool in two different classroom settings. In activity theory the most appropriate unit of analysis is the activity. Since the focus of our Xorro-Q tool is engaging students in simple activities (e.g., answer in one word, select one option) in class, activity theory is particularly well suited to our study of classroom learning. In this theory all purposeful human activities would be a result of interaction among six elements namely subject, object, tool, community, rules and division of labour.

## 4    ACTION RESEARCH METHODOLOGY

For the purpose of this study we have used action research. Creswell et al. (2007) classify action research as practical and participatory. Perhaps the one definition which considers both practical and participatory action research belongs to Rapoport, who defines action research as follows:

> Action research aims to contribute both to the practical concerns of people in an immediate problematic situation and to the goals of social science by joint collaboration within a mutually acceptable ethical framework (Rapoport 1970, p. 499).

Action research is applied in nature which means it starts with a practical problem. Then it attempts to find a solution to the problem. Action research is best applied in educational and organization settings where educators, teachers and practitioners want to reflect on their own practices (Mills 2000).

Generally, action research is designed with some intervention approach in order to change the status quo and to address a concern or solve a specific problem. This research attempts to improve the participation of university students in their classrooms. To address this problem, we have introduced a tool (Xorro-Q) as an intervention which we believe could help increase students' participation in the courses. Quantitative data is collected using the tool. This includes data about student attendance, participation rate of students in the activities and assignments, and their scores in the activities. We also collected qualitative data through open-ended questions from students to get an understanding about their classroom experience. The quantitative and qualitative data have been analysed to help determine whether or not, and, how our tool could help improve students' participation.

## 5 RESEARCH DESIGN

The Xorro-Q tool was used in two undergraduate classroom settings in a New Zealand University (NZU) in the second half of a teaching semester. The courses have been selected randomly. This tool collects quantitative data regarding students' participation in the class. This included number of questions seen by the students and number of questions which were answered by them. In one classroom setting which was a first year computing course, the lecturer employed activity-based teaching techniques with the Xorro-Q tool; while the lecturer in the other classroom setting involving a second year computing course employed Xorro-Q for assessing students subject knowledge using traditional methods. The student enrolment in the first year course was 120 and in the second year course it was 50; however, attendance was about 30% for the first year course and 50% for the second year course. Students were told that there were no marks for in-class participation. The number of questions used in the two classes varied. In the first year course, activity theory elements were applied through the class, where text-based questions were asked and students' responses were displayed to the whole class through word-clouds. In the second year course, traditional teaching pedagogy was employed. The lecturer asked multi-choice and concept type questions at the end of the class to assess the level of understanding among students on the conceptual subject content.

## 6 MAPPING THE RESEARCH DESIGN WITH ACTIVITY THEORY

The activity theory construct has been applied in social sciences and in computing sciences such as human-computer interaction and software development (Fuentes-Fernandez et al. 2007; Georg et al. 2015). The theoretical underpinnings of AT has helped researchers to understand possible mediations between theoretical constructs to achieve desired outcomes. In this study we utilize the social science aspect of activity theory to get recommendations from students on how best we should design a dashboard which appeals to them. In Figure 2, we have aligned the activity theory elements with our study.

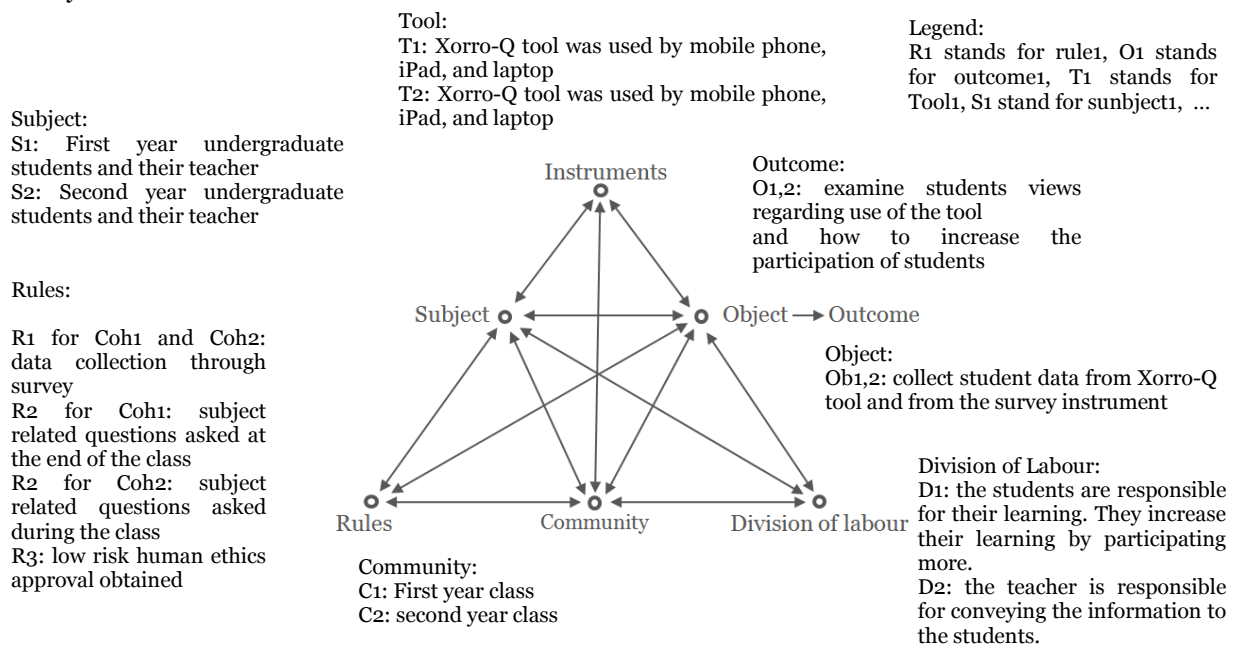

*Figure 2: Mapping activity theory constructs with our study's design*

The seven constructs (subject, instrument, rules, community, division of labour, object/aim and outcome) used in the context of this study are explained next. The *subject* is the teacher considering his/her teaching experience, teaching approach, and the students. The *instrument* is the underlying audience engagement tool Xorro-Q. The *rules* are the expectations of the teachers and the evaluation criteria. The *community* is the class environment which is mediated by rules. In a class, students can discuss their issues with their classmates before they submit their answers. *Division of labour* is related to the students, teachers and their responsibilities in class. The lecturer is responsible for teaching and asking questions through Xorro-Q. The students participate in class activities and are themselves responsible for their learning. For every human activity there is an aim or *object*. The object could be physical or conceptual. The *outcome* would be the result of executing an activity. Our objective is to understand how students considered the use of the tool could help them engage in class and in the process of learning.

## 7   DATA ANALYSIS

We next used Xorro-Q to analyse our data. Xorro-Q has a dashboard to inform teachers and the institution manager about the participation of students and activity of the lecturer in the class. By participation we mean the number of questions seen by the students and the number of answers given by individual students. To find the participation percentage, the two numbers are divided and the percentage is analysed. We categorized the students' participation into four different groups. (1) not attending the class, (2) initiating (participation less than 20%), (3) participating (participation between 20%-80%), (4) engaging (participation more than 80%). Question impression refers to the number of questions which students were exposed to. The tool lists each question as an activity. Therefore, the number of activities run by the lecturer implies the number of questions asked in the classroom.

In the final week of the semester, we used a Student Engagement Questionnaire (SEQ) to capture student perceptions on the use of the audience interaction tool (Xorro-Q) intervention. The online survey (SEQ) was floated to both classes. Overall 15 students from the first year class (i.e., 12.5%) and 18 students from the second year class (i.e., 36%) answered the survey questions. The data collected from the classroom sessions via Xorro Q and the survey questionnaire have been analysed next to gain further understanding on how audience interaction tools can be applied to different classroom environments.

### 7.1   Analysing Students' Participation Data

This section analyses activities of students and lecturers during the class. Figure 3 shows activities which have been run in two classes by two different lecturers. The diagram shows on the number of questions asked in each of these classes. The student participation percentage is calculated based on the number of responses received to the questions asked. Thus in the first year class, 98 questions were asked to a group size of 188 students to which 51% of students responded. In the second year class, 74 questions were asked to a group size of 60 to which 92% of students responded. The size of the circles also shows the size of the class.

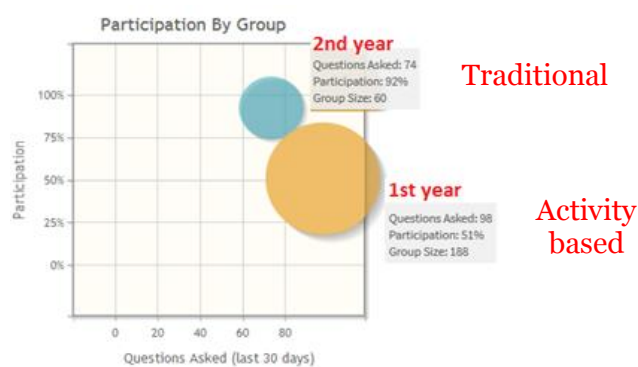

*Figure 3: Group comparison*

The engagement history of the lecturer in the first year and second year courses are shown in Figure 4 and Figure 5. The engagement history refers to how the engagement metrics have been changing over a specified interval of time. The chart describes what proportion of an audience (averaged as determined through the filters) is attending, initiating, participating or engaged.

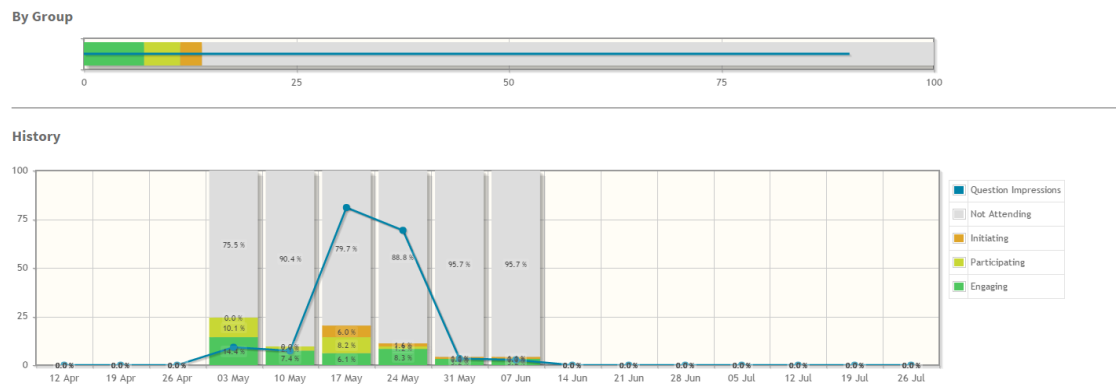

*Figure 4: Engagement history for the first year course*

The figures show how many percentages of students from those students who attended the class were in the initiating, participating, and engaging stage. This is reported by counting the number of participants logged in at the time the question was asked. However, if a participant joins an activity late, then question impressions will not count that participant for the questions asked prior to his joining.

Interestingly Figure 4 and Figure 5 show that the students in the class were engaged more in the second year class (where assessment questions were asked) than the first year class (where discussion questions were asked).

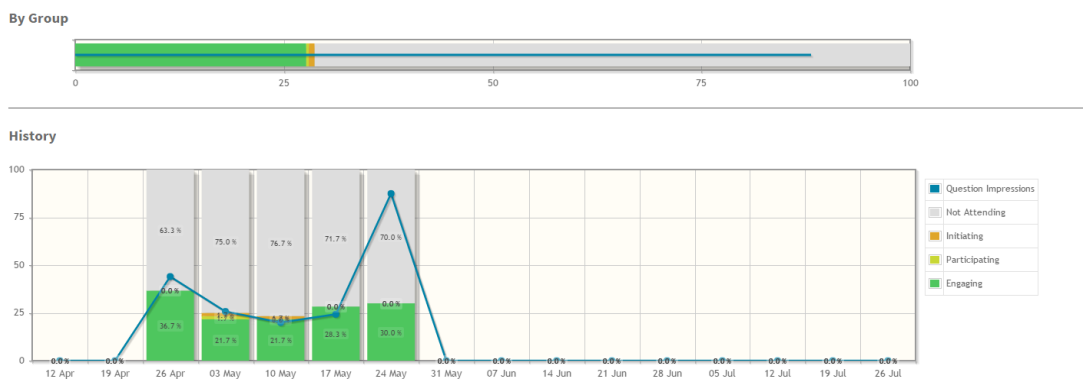

*Figure 5: Engagement history for second year course*

### 7.2  Survey Data Analysis

Statistical analysis in the form of percentage has been used to analyse student participation and satisfaction in regards to the use of Xorro-Q in their classes. Four Likert scale survey questions were used to gauge student satisfaction on the impact of using audience participation tool (Xorro-Q) in their learning process and interest in the course. Students were asked whether raising questions during the classroom sessions through audience interaction tools could help them in their learning process and increase their interest in the course. Students were also asked if asking questions through the tool made them more attentive and engaged in class. Table 1 shows the data from the first year and second year course for these three questions. We realise that the number of responses relative to the class strength is rather low (12.5% for the first year course and 36% for the second year course); however, it gives some indication on how students' view their learning process.

|            | Helped me to learn           || Helped me to increase interest || Made me more attentive and engaged ||
|------------|------------------|-----------|------------------|-----------|------------------|-----------|
|            | C1: 1st yr course | C2: 2nd yr course | C1: 1st yr course | C2: 2nd yr course | C1: 1st yr course | C2: 2nd yr course |
| A lot      | 40%  | 27.77% | 30%  | 22.22% | 70%  | 27.77% |
| Somewhat   | 60%  | 61.11% | 60%  | 55.55% | 20%  | 44.44% |
| A little   | 0    | 11.11% | 10%  | 16.66% | 10%  | 11.11% |
| Not at all | 0    | 0      | 0    | 5.55%  | 0    | 16.66% |

*Table 1: Statistics for students' idea regarding usefulness of the tool*

When the students were asked whether Xorro-Q helped them to learn when they compared their results with other students' results displayed on the board in the classroom, students in class where Xorro-Q tool was continuously used mentioned that the tool helped them when they compared their results with other students' results.

|                            | C1: 1st yr course | C2: 2nd yr course |
|----------------------------|-------------------|-------------------|
| I enjoyed it               | 80%               | 44.44%            |
| It was OK                  | 20%               | 38.89%            |
| I have no opinion as such  | 0%                | 5.55%             |
| It was a waste of time     | 0%                | 11.11%            |

Table 2).

*Table 2: Statistics for students' enjoyment*

Table 3 summarizes general impressions in regards to the asking of questions during the class using Xorro-Q. These findings show that first year students who used the tool continuously in the class enjoy using the tool more compared to the second year students.

|            | C1: 1st yr course | C2: 2nd yr course |
|------------|-------------------|-------------------|
| A lot      | 50%               | 22.22%            |
| Somewhat   | 30%               | 33.33%            |
| A little   | 20%               | 22.22%            |
| Not at all | 0%                | 22.22%            |

*Table 3: Statistics for students' general impression*

It was interesting that in our survey none of the students liked to see the results displayed alongside their names in the class. Students prefer to remain anonymous in public places like classrooms where they can be judged. Although classroom interaction helped them learn as indicated in Table 1, the students were not keen to have their names displayed on the board. These two statements contradict each other. We can only speculate that whenever students are being assessed, they prefer to keep their identity confidential. The next phase of the study will investigate these findings further.

There were also some open-ended questions in our survey instrument. Students' answers to open-ended questions show that students generally have a positive feedback towards using audience interaction tools in classes. Students were asked about their most and least favourite feature in Xorro-Q. According to the students' responses, ease of use, user friendliness, getting immediate feedback during progression in class, and overall classroom interactivity had helped them retain information that they just learnt in the lecture which was the strength of the tool. However, other students indicated that some people were posting irrelevant things to the board which was inappropriate and distracting. Students also did not want their names to be displayed on the board and preferred to remain anonymous.

## 8   DISCUSSION

Questions are raised on data gathered from Xorro-Q regarding engagement history and from SEQ regarding engagement in the class. Table 1 indicates that students are more satisfied when such tools are used continuously to engage them through discussions compared to the time when it was used once to assess their acquired knowledge at the end of the class.

Findings in Table 2 show that in the first year course with activity pedagogy where the tool was used continuously for discussions, 80% of the students who attended classes reported that they enjoyed the ongoing exchanges in the classroom, while 20% of the same students found it satisfactory. In the second year course with traditional pedagogy where the tool was used near the end of the class to assess student learning, 44.5% of the students attending classes reported that they enjoyed answering questions, 39% found it satisfactory, while remaining students either did not like it or were noncommittal. We found that whilst engagement history of the second year class in which traditional teaching methods was employed shows higher level to engagement, the student responses indicate lower level of enjoyment. Similarly in the first year class where activity based teaching methods were employed, while students were less engaged in answering questions through the tool, students enjoyed it more.

Students were asked what their general impression was regarding the asking of questions during the class using Xorro-Q. Our findings show that students enjoyed usage of such tools more for informal discussions rather than for assessment purposes (Table 3). This finding re-affirms the earlier finding that although engaging in assessment was higher, the students do not enjoy being assessed. Students enjoyed more through informal classroom discussions, although some of the students though present did not participate.

## 9   CONCLUSION AND FUTURE WORK

In this study the use of audience interaction tool has been investigated in two undergraduate courses at NZU. Our aim was to examine how using this tool could help students in the process of their learning, and keep them engaged and interested during class. This study drawing on activity theory (Leont'ev 1974) looked at the engagement of students in two undergraduate courses. One of the courses applied a traditional methodology and the other one was activity oriented. Applying the elements of the activity theory as discussed by Engeström (2001) to our classroom settings, we found that audience participation tool has a promising direction for engaging students in the process of teaching and learning. However, engagement in the form of assessment is found to be higher, while engagement in the form of informal discussions may have less student participation, but, is considered more enjoyable. The data collected through our tool showed that the class which adopted an activity-oriented approach rendered better results as far as enjoyment of students is concerned while traditional classes showed higher engagement. The tool has proved to be effective in providing new insights to lecturers, students, and the institution, and has potential to help all the parties involved so that the right decisions are made regarding the efficacy of the teaching and learning, and the evaluation process. However, there are a few limitations in this study. For example, the study has been conducted in the second half of the teaching semester. It would have been better if we could have run the test from the beginning of the course to the end and put students under focus from the start to understand how their level of engagement has changed during the course. Also, the low attendance and low response to the final survey tool are other limitations of this study.

However, the study has shed some light into how audience engagement tools can be used in classrooms when different pedagogies are used. We plan to extend this study with different combinations of pedagogies in different subject areas to understand how such systems affect and engage students in the process of learning. In the process, we hope to tailor certain features of the current on-line audience interaction tools to provide a friendlier space for student-driven learning and for knowledge acquisition and assessment. Future work will extend the experimental design to classes where the teacher and students remain the same, but the teaching pedagogy will vary across traditional, activity-based and team-based settings. This design will be repeated across different undergraduate classes. Further, qualitative data through interviews with teachers and students will be collected to contextualize socio-cultural aspects of teaching and learning practice. It is hoped that this study will offer rich insights on the role of audience interaction tool and have innovative pedagogical implications in teaching and learning.

# Appendix 1

| | Survey Instrument |
|---|---|
| 1 | Did you attend all lectures?<br>• Yes, I attended all the lectures.<br>• I attended majority of the lectures.<br>• I attended a few of the lectures.<br>• No, I could not attend any of the lectures. |
| 2. | What was your general impression regarding asking of questions during the class using Xorro-Q?<br>• I enjoyed it.<br>• It was OK<br>• I have no opinion as such.<br>• It was a waste of time. |
| 3 | How easy was it to use the software tool "Xorro-Q"?<br>• Very easy<br>• I worked it out<br>• Not intuitive<br>• Very hard |
| 4 | Do you think that by using the technique of asking questions through Xorro-Q, we can help you in the process of learning?<br>• A lot<br>• Somewhat<br>• A little<br>• Not at all |

| 5 | Do you think that by asking questions (with Xorro-Q), could highlight the gaps in your knowledge?<br>• A lot<br>• Somewhat<br>• A little<br>• Not at all |
|---|---|
| 6 | Do you think that by asking questions (with Xorro-Q), we can increase your interest in the course?<br>• A lot<br>• Somewhat<br>• A little<br>• Not at all |
| 7 | Do you think answering questions (using Xorro-Q) made you more attentive and remain engaged in class?<br>• A lot<br>• Some what<br>• A little<br>• Not at all |
| 8 | Did it help you to learn when you compared your result with other students' result when it was shown on the board?<br>• A lot<br>• Somewhat<br>• A little<br>• Not at all |
| 9 | How does it feel to see your answers displayed for everyone to see them?<br>(Select as many as apply)<br>• I enjoy it<br>• It motivates me<br>• It makes me more attentive<br>• I find it scary |
| 10 | How would you like to see yourself displayed on the board?<br>• By name<br>• By ID<br>• Anonymous<br>• I do not care, either way |
| 11 | Do you think that by answering questions through Xorro-Q, it may have helped you in preparation for the exam?<br>• Yes<br>• No<br>• I don't know |
| 12 | How do you like to study for the exam individually or in a group?<br>• Prefer individual study<br>• Prefer group study |
| 13 | What sort of mobile device did you use to connect to Xorro website? |
| 14 | What did you like about the tool (Xorro-Q) that we used in the class? |
| 15 | What did you NOT like about the tool (Xorro-Q) that we used in the class? |
| 16 | Can you suggest any improvements to the tool (Xorro-Q) we used in the class? |
| 17 | What would you have liked to be done differently in the class to engage you more? |